# Growth and Properties of Dislocated Two-dimensional Layered Materials

Rui Chen[1,2], Jinhua Cao[1,2], Stephen Gee[1,2], Yin Liu[1,2,3]* and Jie Yao[1,2]

1 Department of Materials Science and Engineering, University of California, Berkeley, Berkeley, CA, USA.

2 Materials Sciences Division, Lawrence Berkeley National Laboratory, Berkeley, CA, USA.

3 Present address: Department of Materials Science and Engineering, Stanford University, Stanford, CA, USA.

## ABSTRACT

*Two-dimensional (2D) layered materials hosting dislocations have attracted considerable research attention in recent years. In particular, screw dislocations can result in a spiral topology and an interlayer twist in the layered materials, significantly impacting the stacking order and symmetry of the layers. Moreover, the dislocations with large strain and heavily distorted atomic registry can result in a local modification of the structures around the dislocation. The dislocations thus provide a useful route to engineering optical, electrical, thermal, mechanical and catalytic properties of the 2D layered materials, which show great potential to bring new functionalities. This article presents a comprehensive review of the experimental and theoretical progress on the growth and properties of the dislocated 2D layered materials. It also offers an outlook on the future works in this promising research field.*

*e-mail: yin6@stanford.edu

# 1. INTRODUCTION

A dislocation is a linear crystallographic defect extensively present in crystals[1]. Dislocations play an important role in the crystal growth, which provide active sites for the nucleation. Especially, screw dislocations drive the spiral growth of crystals through the advancement of steps at kinks formed by the intersection of a screw dislocation with the surface[2]. The dislocation driven growth has a profound effect on the growth rate and is a ubiquitous mechanism for the crystal growth in low supersaturation conditions.

Dislocations can significantly affect the properties of materials. Understanding dislocation behaviors is crucial for the interpretation of the mechanical properties and plastic deformation of solids[3]. Additionally, dislocations can significantly affect optical, electronic and optoelectronic properties of materials. The effect of dislocations on these physical properties is manifold. In many optoelectronic devices, dislocations scatter electrons and phonons, and act as non-radiative recombination centers, deteriorating the performance of the devices. Besides, the heavily distorted atomic registries and the large strain of the dislocations can give rise to exceptional properties that are not present in the host materials. Dislocations have been judiciously engineered to bring new functionalities into materials. For example, dislocations in insulating materials can act as conducting channels. The conductive dislocation filaments have provided an essential mechanism for the resistive switching technologies[4]. It has also been suggested that dislocations can generate exotic topological electronic states for dissipationless electron conduction in topological materials[5]. Dislocations are intentionally introduced into thermoelectric materials to boost the efficiency of the heat-to-electricity conversion[6]. Dislocations can be utilized as a useful tool to tune the magnetic and ferroelectric phase transitions. Dislocations in paramagnetic materials can enable ferromagnetism and dislocations in paraelectric materials can generate ferroelectricity[7,8].

In recent years, tremendous research efforts have been concentrated on two-dimensional (2D) van der Waals (vdW) layered materials and dislocations in such systems[9-11]. In particular, screw dislocations drive the growth of nanoplates of various 2D layered materials with a spiral topology[12-14]. More recently, it has been reported that screw dislocations lead to the growth of helical layered crystals with a tunable Eshelby twist[15,16]. Screw dislocations not only influence the morphologies of 2D materials, but also markedly modulate the symmetry and stacking order of layers, providing an intriguing route to modifying the optoelectronic and valleytronic properties of 2D materials. Screw dislocations in the layered materials have a unique topology. The dislocation core spirally threads the atomic layers, providing a helical path for the electron and phonon transport. The high strain and highly disordered atomic registry can result in a local modification of the chemistry and properties around the dislocation core. Many studies have suggested intriguing properties of dislocated 2D layered materials, including enhanced nonlinear optical properties[12,13], valley polarization[14], electrical conductance[17,18], and catalyst performance[18,19]. Despite a large body of research articles on the dislocated 2D materials, there exists no review article devoted to the topic thus far to the best of our knowledge.

Here, we provide a comprehensive review of the recent experimental and theoretical developments on the dislocated 2D layered materials. This article primarily focuses on layered materials with screw dislocations. In section 2, we will focus on the screw dislocation driven (SDD) growth of layered materials. In section 3, we will discuss the modification of the physical properties of layered materials introduced by screw dislocations. When 2D materials are thinned down to the monolayer and bilayer regime, dimensional confinement is applied to the Burgers vector of dislocations; the dislocations under such cases can behave differently. In section 4, we will provide a brief survey of dislocations in monolayer and bilayer 2D materials. In section 5, we will envision two main promising directions to be pursued based on dislocated 2D layered materials. Finally, in section 6, we will summarize the crucial aspects in this review article.

## 2. DISLOCATION DRIVEN GROWTH OF LAYERED MATERIALS

### 2.1 Dislocation driven growth of layered spiral plates

In conventional three-dimensional (3D) crystal growth, the formation of a 2D nucleus on a substrate is thermodynamically suppressed by the creation of new interfaces. Hence, incredibly high supersaturations are required to overcome this high free energy barrier and allow for the growth of crystals. Here, as the driving force of the crystal growth, the supersaturation $\sigma$ is defined by $\sigma = \ln(c/c_0)$, where $c$ and $c_0$ are the concentrations of precursors and equilibrium, respectively. However, experiments have shown that a large growth rate can be achieved at low levels of supersaturation when accompanied by dislocated surface steps[20]. This phenomenon is explained by a new growth mode described by Burton-Cabrera-Frank (BCF) crystal growth theory[20]. In practice, crystals can harbor screw dislocations which generate atomically stepped surfaces (FIG. 1a) and thus substantially decrease the required high supersaturation (approximately from 50% to 1% or less[21]) and high temperature to create an initial nucleus during the growth[20]. The intersection of a screw dislocation with a crystal surface holds great potential for triggering the winding SDD growth of not only 3D but also 2D materials, such as layered spiral nanoplates (FIG. 1b,c)[12,13,20-23]. For the introduction of screw dislocations in vdW materials such as layered chalcogenides, the misalignment of two surface steps must occur under the following conditions: (a) the growth rates of two adjacent steps differ from each other; (b) the existence of chalcogen-rich regions yields a linear accumulation of metal vacancies; (c) a high growth rate lifts the relatively faster-growing step over the slow one (FIG. 1d)[24]. Also, exterior contributions, such as defective substrates and microscopic inclusions[22,25-27], can be highly crucial for the initial introduction of screw dislocations.

The core of spiral growth is the creation of anisotropic growing kinetics through dislocations. Consequently, the growth process is heavily dependent on the local supersaturation state, which determines the dislocation driven growth, layer-by-layer (LBL) growth, and dendritic growth modes. Figure 1e depicts that the dislocation driven growth dominates in the low supersaturation region. Whereas an increasingly higher supersaturation suppresses the dislocation driven mode, and progressively favors LBL and dendritic growth modes[22]. Overall, supersaturation tailors the growth rate, the symmetry of the growing samples, and the crystal size among other properties[23].

There are three main strategies for the spiral growth of layered nanoplates: vapor, solution and epitaxy techniques. Vapor methods, such as chemical vapor deposition (CVD)[12,13,17,18,23,28-33] and chemical vapor transport (CVT)[14], are the most widely utilized owing to their high growth rates and good reproducibility. The growth of spiral structures using vapor methods is very sensitive to the local supersaturation of the adatoms near the dislocation ridge which depends on reaction temperature, the mass-transfer velocity and the precursor loss for the prior deposition (namely, the temperature distribution in the furnace)[23]. In this case, diverse spiral structures can be produced in a controllable manner. For example, in transition metal dichalcogenides (TMDCs, such as $MoS_2$, $WS_2$, $MoSe_2$, and $WSe_2$), the as-synthesized structures span from single to double and further to multiple spiral patterns, which are spatially distributed in the growth region because of the temperature gradient[23]. The aqueous solution method also serves as a promising alternative to the vapor ones[34-36], in light of its great potential to produce free-standing spiral plates. Similarly, the supersaturation is governed by the amount of precursors, the pH value, the reductant concentration, among other parameters[34]. Molecular beam epitaxy (MBE) methods also have their advantages for producing spiral layered materials with high crystalline quality due to their precisely controlled film uniformity, thickness and defect level[24]. Also, the density of spiral ridges can be effectively tuned by the beam flux. The higher the flux is, the finer the spiral thread will be[24,37]. This ability to precisely tailor the growth makes MBE particularly valuable for applications in modern semiconductor electronics.

Quite interestingly, the growth of layered spirals also depends on several other factors. For instance, the step height of dislocation ridge influences the number of screws harbored in the crystals, determining the single or multiple spiral domain growth modes. In addition, there exists high flexibility between any two screw dislocations in one spiral plate: no spatial overlap is required; two sets of spiral ridges can be arbitrarily twisted[24]. As a result, the abundant categories of layered spirals hosting varying structural symmetries yield the ampler and more interesting physics in vdW materials, together with great promises for the practical applications.

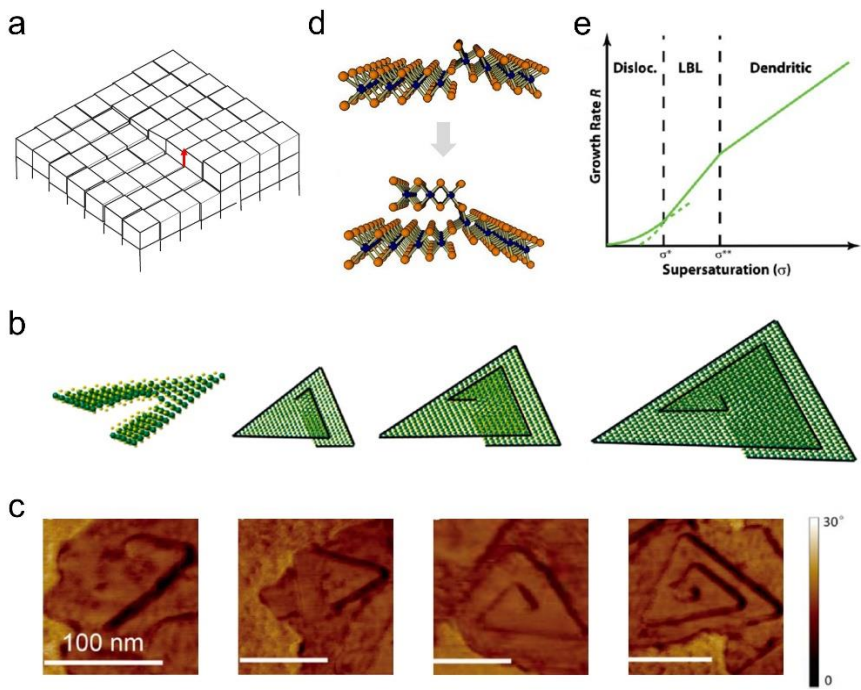

Figure 1 | **SDD growth of spiral nanoplates. a**, Schematic illustration of a screw dislocation in the crystalline solids. The red arrow denotes the Burgers vector. **b, c**, Schematic diagrams (**b**) and atomic force microscopy (AFM, **c**) of the growing spiral nanoplates driven by a propagating screw dislocation. Here, the AFM phase signals are used to investigate the morphology of $MoS_2$ spirals Scale bar: 100 nm. **d**, Two adjacent domains of $WSe_2$ intersect with each other, lifting the boundary and then generating a screw dislocation. **e**, Growth rate as a function of supersaturation, showing dislocation driven, LBL and dendritic growth modes. Figures reprinted with permission from [21] (**a**), [12] (**b, c**), [38] (**d**), and [22] (**e**).

## 2.2 Dislocation driven growth of twisted vdW nanowires

Twisted 2D materials have recently attracted considerable research attention, leading to the fast growing research field of 'twistronics'[39-44]. The twist between the atomic layers can create 'Moiré patterns' with long-range periodicity, resulting in unconventional properties such as superconductivity, ferromagnetism and Moiré trapped excitons[40,42,43]. The twist angle has been shown to be a crucial tuning knob for controlling these properties in twisted 2D materials.

In contrast to the conventional transfer-stacking method used to create twisted 2D materials, it has recently been shown that twisted vdW structures can be synthesized with a tunable twisting topology using a bottom-up synthesis method[15]. The twist originates from a screw dislocation and its induced crystallographic twist, a so-called Eshelby twist. J. D. Eshelby predicted that an axial screw dislocation at the center of a thin whisker can result in a twist in the whisker[45,46]. As a result, the elastic energy associated with the screw dislocation is reduced. In Eshelby's theory, the twist rate is defined as

$$\alpha = \kappa \frac{b}{A},$$

where $b$ is the Burgers vector of the screw dislocation, $\kappa$ is the geometrical prefactor that depends on the shape of the cross section and $A$ is the cross sectional area of the whisker. The Eshelby twist has been identified as an intriguing mechanism for the growth of various twisted nanowires including lead sulfide, lead selenide, zinc oxide, cadmium selenide, cuprous oxide and indium phosphide[47-51].

Even though screw dislocations in layered vdW materials are well known[12-14], the Eshelby twist has not been explored in the nanoplates of layered materials since screw dislocation in those materials have a relatively large lateral size (larger than a few micrometers), resulting in a negligible twist. In contrast, dislocated vdW nanowires displaying compressed cross sectional areas open up new opportunities. It has been demonstrated that screw dislocations can drive the formation of twisted germanium sulfide (GeS) vdW crystals on scales ranging from nanoscale to microscale[15,16]. In the synthesis, GeS nanowires were first grown with the growth direction along the vdW stacking direction. Introducing a screw dislocation into such nanowires naturally leads to the Eshelby twist between the successive layers (FIG. 2c,d). These nanowires possess continuous twists in which the total twist rates are defined by the radii of the nanowires, consistent with Eshelby's theory (FIG. 2e). Further radial growth of those twisted nanowires that are fixed to the substrate leads to an increase in elastic energy, as the total twist rate is fixed by the substrate (FIG. 2a). The stored elastic energy can be reduced by accommodating the fixed twist rate in a series of discrete jumps in the twisting profile. This yields mesoscale twisted structures consisting of helical assemblies of vdW layers demarcated by atomically sharp interfaces with a range of twist angles (FIG. 2g-j).

The twisting morphology of the structures and the interlayer twist angles are tunable. The twisting morphology gradually transitions from initial continuous twisting to intermediate twisting (consisting of both continuous twisting between the twist boundaries and discrete twisting at the boundaries) and eventually to discrete twisting with increasing radial size (FIG. 2a,b). This allows for the control of twisting profile and angles at twist interfaces by controlling the radial growth of the structure.

The twist rates and periods of the structures are determined by the radii of the dislocated nanowires that are defined by the size of the Au-Ge alloy droplets catalyzing the vapor-liquid-solid (VLS) process. Through introducing germanium selenide (GeSe) into the growth, the droplet size was chemically tuned and thereby the twist rate and period of the twisted structures[52]. The chemical modulation demonstrates good potential to tailor the twist rate and period of helical vdW crystals, enabling a new degree of freedom to modulate optical, electrical, thermal, mechanical and catalytic properties.

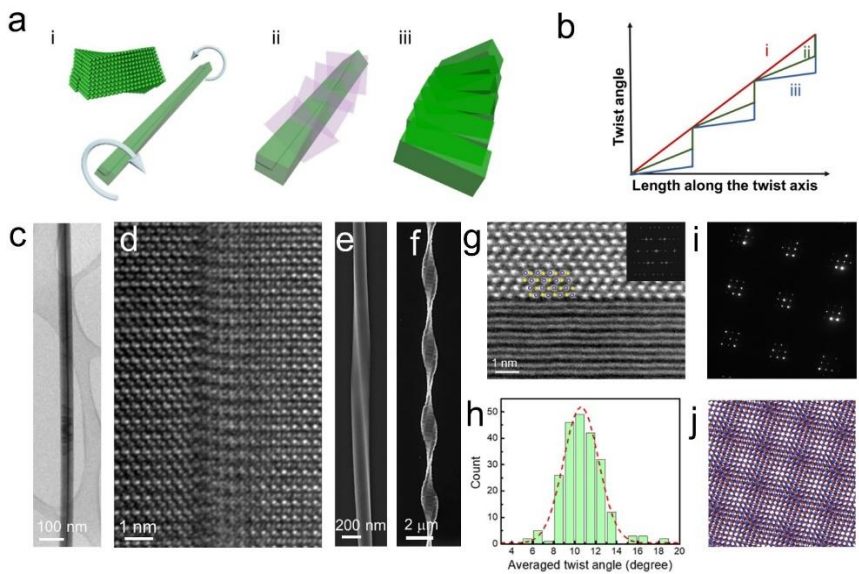

Figure 2 | **SDD growth of twisted GeS nanowires. a**, Schematics showing the evolution of the twisting morphology with increasing radial size of the structure. (i) The growth of nanowire with Eshelby twist. The inset shows a schematic of the atomic structure of the Eshelby twist; (ii) interlayer slip that forms the twist grain boundary; (iii) further radial growth giving rise to the discretely twisting morphology. **b**, Schematic diagram showing the twisting profile transitions from a continuous twist to a discrete twist corresponding to states (i-iii) from (**a**). **c**, TEM image of a nanowire with Eshelby twist. The arrow denotes the screw dislocation in the middle of the nanowire. **d**, High resolution TEM image of the dislocation. **e**, Scanning electron microscopy (SEM) image of a twisted nanowire adhering to a substrate, showing a continuous twisting profile. **f**, SEM image of a microscale twisted crystal showing a discrete twist formed by helically stacking nanoplates. **g**, High-angle annular dark field scanning transmission electron microscopy (HAADF-STEM) image showing the atomically sharp interface between two nanoplates. The inset shows the fast Fourier transform (FFT) pattern of the upper crystal on the [010] zone axis. **h**, Distribution of twist angles in the microscale crystals. **i,j**, Double diffraction pattern acquired from a twist interface with a twist angle of 7.5 degree (**i**) and its corresponding Moiré pattern (**j**). Figures reprinted with permission from [15].

## 3. NOVEL PROPERTIES IN THE DISLOCATED LAYERED MATERIALS

### 3.1 Nonlinear optical properties

The unique crystal symmetry of spiral layered materials plays a critical role in generating novel optical phenomena. A broken inversion symmetry in the layer stacking allows for intriguing nonlinear optical effects, especially the second-harmonic generation (SHG, FIG. 3a). For instance, in the TMDC system, three types of spiral structures have been extensively studied: triangle, hexagon and their mixture. These diverse spiral modes are tailored by the local supersaturation during the SDD growth; with lowering supersaturation, triangular, hexagonal and multiple spirals are favored successively[23]. As shown in FIG. 3b, triangular TMDC plates are formed by 'AA' layer stacking, where the adjacent two layers show 0º rotation and break the inversion symmetry ($D_{3h}^1$ group). Thus, triangular TMDC spirals display strong SHG signals[12,13,23,33]. In contrast with monolayer TMDC, the SHG intensities from the TMDC spirals are substantially enhanced by the large constructive interference length[12,30,33] and the strong interband Berry connection[33]. On the contrary, hexagonal TMDC plates, which are formed by 'AB' layer stacking with an inversion center ($D_{3d}^1$ group), yield vanishing SHG (FIG. 3b)[13]. Also, no SHG is observed in a pseudo-2H assembly consisting of double triangular spiral patterns with a 60º rotation angle[13,23]. The combination of triangular and hexagonal spirals gives rise to a mixed structural phase where the SHG is strongly suppressed and spatially non-uniform[13]. Nonetheless, an enhanced SHG has been reported in the 5º twisted screw structures[30], double triangular spiral patterns with a 120º included angle[23] and five triangular spiral patterns[23], to name just a few. Indeed, screw dislocation plays a critical role in engineering the layer stacking symmetry and nonlinear optical properties in layered materials. Such a strong SHG response in the TMDC spirals significantly relaxes the stringent requirements in the production of atomically thin odd-layer 2D materials[53], paving way for the emergent nonlinear optical applications based on spiral nanoplates. Besides SHG, a large third-harmonic generation effect occurs in the CVD-synthesized $WS_2$ spiral structures, originating from a profound third-order nonlinear susceptibility due to dislocations[30].

## 3.2 Valley physics

Valley-based electronics and optoelectronics have been researched widely in the last decade, such as valley Hall effect, monolayer valley excitons, valley depolarization and interlayer excitons in bilayer heterostructures[54]. However, such exceptional valley physics requires their inversion symmetry breaking and thus is generally existing in the monolayer TMDCs[55]. Non-centrosymmetric TMDC spirals have been shown to enable new access to the valley and spin degrees of freedom. In the meantime, the fascinating interplay between valley and spin relies on the strong spin-orbital coupling. The first observation of valley-contrasting spin splitting in the dislocated $3R-MoS_2$ crystals (non-centrosymmetric triangular spirals which are similar to the 'AA' stacking if the interlayer translation is ignored) was obtained by means of spin- and angle-resolved photoemission spectroscopy (SARPES, FIG. 3c,d)[14]. The considerable energy splitting in the spirals, ~0.14 eV, is comparable to that in monolayer TMDCs[55-58]. Also, photoluminescence (PL) circular dichroism suggests a large and thickness-independent valley polarization (~50-60%), induced by highly suppressed interlayer/intravalley hopping in the 3R stacking (FIG. 3e)[14]. This proposed relaxation model was experimentally verified in the triangular $WS_2$ spiral via a distinct linear polarization in the PL spectra[31]. Additionally, a remarkably higher degree of valley polarization (~94±4%) was achieved in tungsten dichalcogenides using lower measurement temperatures and near-resonant optical excitation to reduce interlayer scattering[31]. These exotic valley/spin phenomena make TMDC spirals an intriguing alternative to the single layers for the more in-depth study of the valley/spin correlation and manipulation.

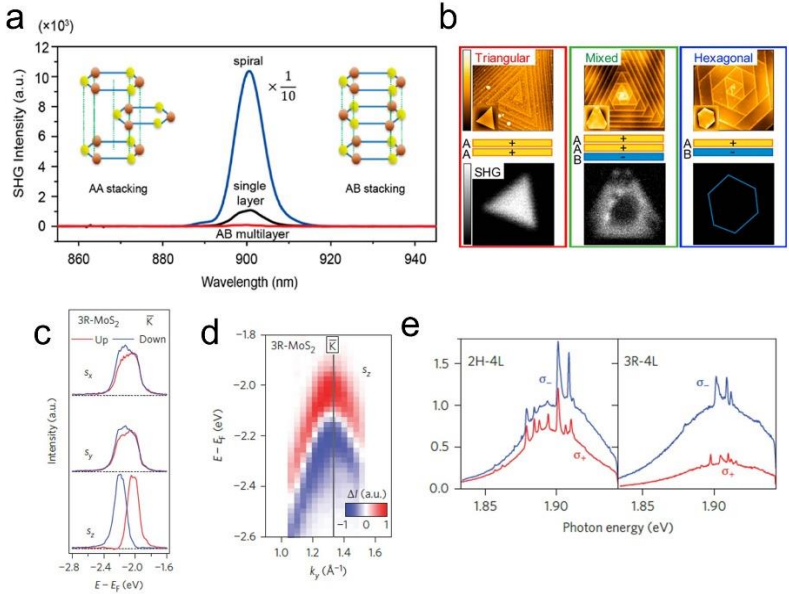

Figure 3 | **Exotic optical properties. a**, Nonlinear optical SHG signal from CVD-grown monolayer, exfoliated AB stacked multilayer and SDD grown AA stacked spiral of $MoS_2$. Insets show the side view of AA and AB stacked multilayer lattice, respectively. **b**, AFM and SHG images of representative $WSe_2$ nanoplates with different dislocation spiral behaviors. **c**, Spin-resolved energy distribution curves for SDD grown 3R-$MoS_2$ obtained from SARPES. **d**, Valley-contrasting spin splitting of SDD grown 3R-$MoS_2$ imaged by SARPES. **e**, PL circular dichroism of SDD grown 3R-$MoS_2$ and 2H-$MoS_2$. Figures reprinted with permission from [12] (**a**), [13] (**b**) and [14] (**c-e**).

## 3.3 Phonon behavior

The ways of the layer stacking can affect the phonon modes in the spiral structures. As a result, Raman spectroscopy has been used to explore new lattice vibrational features, particularly in the spiral TMDC systems. In the high frequency (HF) regime, the Raman features of hexagonal TMDC spirals resemble those of 2H-bulk, indicating almost the same lattice vibrations. However, triangular TMDC spirals exhibit a tiny splitting between the $A_{1g}$ and $E^1_{2g}$ phonon modes as well as an enhanced intensity, indicative of non-degenerate out-of-plane and in-plane lattice vibrations[13]. Also, the interlayer mechanical coupling in the triangular TMDC spirals can trigger a softer $E^1_{2g}$ mode and a stiffer $A_{1g}$ mode compared with their monolayer counterparts[12,59] which agrees well with the thickness-dependent trend observed in the 2H-stacked multilayer TMDCs[60]. On the other hand, the low frequency (LF) Raman, which is susceptible to the layer stacking, offers abundant signatures of interlayer breathing and shear in the layered spirals[13,28,61]. A similar layer stacking symmetry was identified among the hexagonal 2H spiral, the double triangular (pseudo-2H) spiral and the 2H-bulk forms of $WSe_2$ with similar peak positions of shear modes. Meanwhile, a much weaker interlayer interaction was observed in the triangular phase which drives the redshift of the shear mode[13]. Considering phonons serve as efficient heat carriers, such unique phonon features in the dislocated 2D materials can make a fundamental difference to the thermal conductivity in the crystals: a notably enhanced axial thermal conductivity has been theoretically predicted in the graphene helicoid compared with non-spiral graphite (FIG. 4a). This enhancement can be interpreted by the divergent in-plane thermal conductivity of graphene, significantly stronger interlayer overlap and potentially a small compressive strain[62]. Further efforts will be required to dig into more complicated phonon systems, such as the chirality-engineered phonon modes, surface phonon modes, etc., together with tailored thermal management and the coupling with electromagnetic excitations to form surface phonon polaritons.

### 3.4 Bandgap tuning

Bandgap tuning has long been the focus of research interests in the 2D semiconductors which enables the engineering of electronic and optoelectronic properties. It is well known that 2D TMDCs adopt an indirect-to-direct-bandgap crossover when the layer number goes down to one[63]. Interestingly, these exciton modes can also be effectively modulated by the unique spiral structures. It is worth noting that triangular, hexagonal and mixed TMDC spirals can all harbor the A exciton mode, rather than the indirect bandgap emission in the 2H-bulk[13]. Also, a redshift of peak position resolved by PL in the triangular spirals manifests a suppressed bandgap size, in contrast with both monolayer[12,64] and hexagonal spiral[13] cases. More interestingly, the existence of dangling bonds and high local stress in the dislocation center can remarkably reduce the bandgap size, even to zero, yielding an exotic phase transition from insulating to metallic states[32]. In addition to the TMDC systems, not only a more significant bandgap reduction (up to 0.6 eV) but also a shorter carrier lifetime was reported in the SDD bismuth oxychloride (BiOCl) spiral nanosheets because of the Moiré-pattern-induced stronger interlayer couplings[19]. Recently, twisted nanowires of 2D layered materials also show a continuous modulation of the bandgap along the nanowires due to the spatially varying strain (FIG. 4b)[15].

### 3.5 Electrical properties

Inefficient electron tunneling through the vdW layers of 2D semiconductors results in poor vertical electrical conduction which may impede their state-of-the-art device applications. However, spiral 2D layered materials can overcome such challenges since the threading dislocations can connect the transverse transport in each atomic layer, thus yielding a considerable vertical transport. Such excellent vertical conductance was experimentally demonstrated in TMDC spiral pyramids (FIG. 4c,d)[17,18]. Interestingly, theoretical calculations also indicated that the superior vertical conductance in a graphene nanosolenoid can serve as a remarkably strong and confined magnetic inductor, owing to the unique helicoid topology[65]. In addition, such helically twisted structures provide a promising playground for studying strong correlation physics. First, Weyl nodes were predicted in chiral twisted graphene owing to the broken inversion symmetry, with the type-I or type-II Weyl phases engineered by the magic twist angle. Second, a twisted Weyl semimetal exhibits a strong correlation between Moiré patterns and Weyl physics[66]. Notably, the spiral morphology is able to promote optoelectronic performance. Figure 4e shows a significantly faster photoresponse in the microsecond regime. For comparison, the response time is at the millisecond level for photodetectors based on few-layer $SnSe_2$ sheets[67,68]. Meanwhile, an extraordinary photoresponsibility has been exhibited in the spiral $SnSe_2$ crystals[29], holding great promises for the high-speed, high-performance and high-efficiency photodetectors.

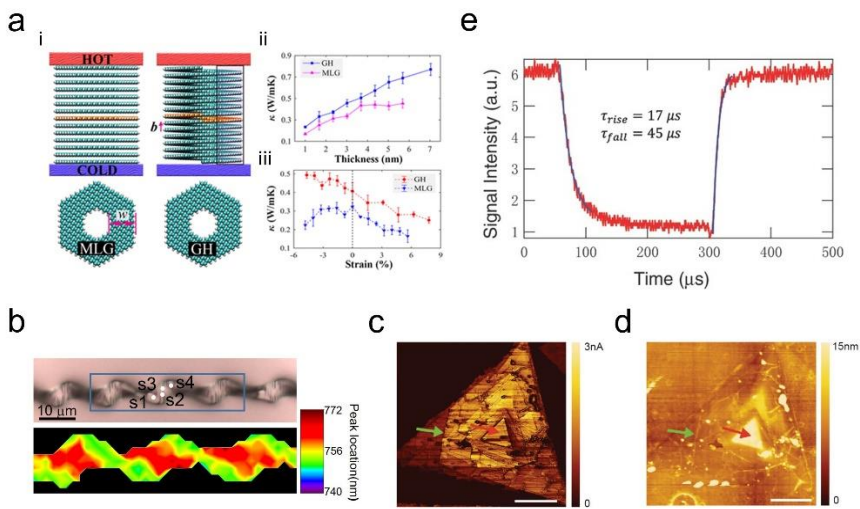

Figure 4 | **Exceptional thermal, electronic, and optoelectronic properties. a**, (i) Schematics of heat conductors made of multilayer graphene (MLG) and graphene helicoid (GH) constructed from screw dislocations. (ii) and (iii) Heat conductivities of GH heat conductors are higher than the MLG counterparts for varying thickness and strain. **b**, Optical image of a twisted GeS nanowire (top) and the corresponding PL mapping of varying peak wavelength positions (bottom). Bandgap size is effectively tuned by the local strain. The blue frame denotes the selected mapping area. **c**, Current conductive AFM (c-AFM) image of an SDD grown $MoS_2$ spiral pyramid ($V_{bias}$ = 2 V). **d**, Tomography c-AFM image of the same $MoS_2$ spiral pyramid. **e**, Rise and fall time of the dislocated $SnSe_2$ photodetector within the microsecond regime. Figures reprinted with permission from [62] (**a**), [15] (**b**), [17] (**c, d**) and [29] (**e**).

## 3.6 Mechanical properties

The addition of dislocations into 2D layered materials can lead to interesting improvements to mechanical performance. Tobermorite, a naturally occurring layered calcium silicate-hydrate material, has attracted immense attention for its engineering applications as a cement. According to molecular dynamics simulations, first, a single screw dislocation, either perpendicular or parallel to the layers, drastically increases the shear strength and toughness of the tobermorite. Second, the atomic distortion around the screw dislocation is believed to result in dislocation climbing, thus preventing the dislocation glide along the core. This leads to improvement in the mechanical behavior, and transforms tobermorite from a brittle to a ductile solid (FIG. 5a)[69]. The spiral symmetry also acts as an ideal system to study exotic mechanical phase transitions. As shown in FIG. 5b, a normal-type $MoS_2$ nanospring generated by a screw dislocation was theoretically expected to undergo a structural transition from a homogeneous to inhomogeneous morphology driven by its unique interlayer vdW interactions[70]. Remarkably, the deformation of the inhomogeneous structures displays a notable deviation from Hooke's law as new surfaces are created during this transition which influence the total strain energy[70]. This inhomogeneous structural transition is expected to be universal in the nanosprings based on most 2D material families[70], owing to similar surface energies[71,72]. It is important for future work to experimentally examine such exotic mechanical behaviors, the control of which will fuel opportunities for a wide range of applications, such as nanomachines, nanoscale mechanical systems, and nanorobots.

**3.7 Catalytic performance**

Catalytic properties can be significantly improved by dislocations in the layered materials. Among these spirals, the recently discovered TMDC spirals can serve as promising catalysts for the hydrogen evolution reaction with exceptionally improved efficiency, exhibiting a smaller Tafel slope and lower onset potential (FIG. 5c,d) in stark contrast with the monolayer and multilayer phases with 2H stacking[18]. This superior catalytic performance is attributed to the strongly coupled edge sites by the dislocation lines[22]. Meanwhile, the formation of Moiré superlattices (MSLs) can also modulate the optoelectronic properties in the layered spirals, thus improving the photocatalytic performance. Compared with conventional bulk BiOCl, MSLs in the spiral nanosheets display a remarkably higher photocatalytic efficiency in the visible-light regime (FIG. 5e). (For comparison, when the degradation efficiency of Rhodamine B approaches 99% in MSLs, the efficiency still stays in 73% and 62% in the bulk reference, under UV-Vis and visible irradiations, respectively). Such enhanced photocatalytic efficiency is attributed to the MSL-tailored optoelectronic properties, including the shrinking bandgap into the visible region and photon-generated charge separation[19]. As discussed above, the spiral growth of vdW materials provides a fascinating playground to research defect-engineered catalytic properties for both science and technology advances.

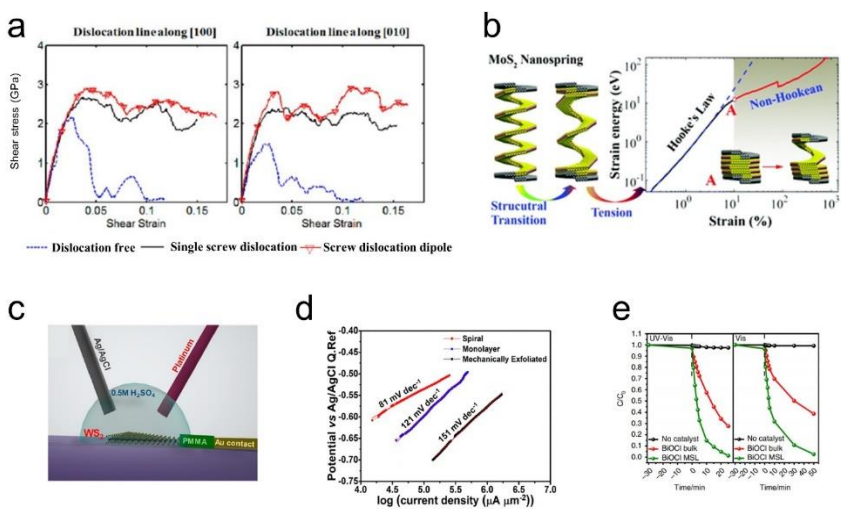

Figure 5 | **Intriguing mechanical and catalytic properties. a**, Calculated stress−strain relationships for tobermorite with a single screw dislocation, a dipole screw dislocation, and no dislocation. **b**, Schematic of MoS$_2$ nanospring and calculated stress-strain relation with the breakdown of Hooke's law. **c**, Schematics of the electrochemical microcell assembly for electrocatalytic measurements. **d**, Tafel plots obtained from SDD grown spiral, grown monolayer and exfoliated thick flakes of WS$_2$. **e**, Photocatalytic degradation rate of Rhodamine B versus time without catalyst and with BiOCl bulk, SDD grown BiOCl MSL nanosheets as catalysts. Negative time denotes dark condition before light irradiation. Figures reprinted with permission from [69] (**a**), [70] (**b**), [18] (**c, d**) and [19] (**e**).

## 4. DISLOCATIONS IN THE MONOLAYER AND BILAYER 2D MATERIALS

The dislocations in monolayer 2D materials, such as monolayer graphene, hexagonal boron nitride (h-BN) and TMDCs, have also been extensively studied with a variety of experimental approaches[73-79] and first principles calculations[80-82]. Different from the dislocations in 3D bulk materials[1], the Burgers vector of dislocations in the monolayer 2D materials are constrained in the material plane[80]. Due to this distinct dimensional confinement, dislocations are now degraded from line defects into point defects. For instance, dislocations in graphene consist of pairs of pentagons and heptagons[78,79]. This construction preserves the coordination environment of all carbon atoms. Although monolayer h-BN and TMDCs share a quite similar honeycomb structure to that of graphene, one significant difference is the polarity of their covalent bonds. This polar nature enables the existence of competing dislocation structures in these materials[81,82]. Odd-membered rings present in dislocations in monolayer h-BN and TMDCs break the altering order of two atomic species, which results in homoelemental bonding and increased formation energy. In this case, edge-sharing pairs of four- and eight-membered rings will also be possible although the elastic contributions to the formation energies are larger when compared with the pentagon-heptagon pair case. In contrast with the monolayer cases, when there are two layers of 2D materials, the stacking manner of the two layers provides a new degree of freedom, partial dislocations separating regions with different stacking manners are also studied[83-85]. Due to the absence of stacking-fault energy, dense dislocation network is present in CVD grown bilayer graphene[84]. A striking linear magnetoresistance was experimentally demonstrated in this system owing to its internal mosaic tiling[86], which provides an ideal two dimensional platform for the study of inhomogeneous conductors.

## 5. OUTLOOK

We have summarized the SDD growth of 2D layered materials, along with the recently reported intriguing properties. However, the field remains in its infant stage, and the whole community are still lacking a comprehensive understanding of the SDD growth mechanisms, such as the nucleation, substrate engineering, introduction and propagation of dislocations. Knowledge of the growth dynamics will enable a better controlled symmetry, chirality, morphology and so on, which paves way for the future investigations. With a controllable twisted morphology driven by dislocations, it is fundamentally exciting to approach the twisted vdW crystals with controllable twisting angles and the extension to other 2D vdW material families (such as graphene, h-BN, black phosphorus and layered oxides) or complex heterostructures. The developments of materials growth will potentially overcome the challenges in manual stacking techniques, and open a new avenue to study the Moiré-pattern-engineered flat-band superconductivity[40,87], ferromagnetic interactions[43,88], excitons[89,90], phonon polaritons[91], to name just a few.

Besides growth, investigations into dislocation-induced peculiar properties are of great importance. First, the axial vdW nanowires and nanoplates with a tunable twist may have a large and tunable chiroptical response. Especially, the twisted nanowires could support and waveguide unique optical modes with the optical spin and orbital angular momentum, which can be used for optical communication and imaging[92,93]. Second, experimental efforts are strongly demanded to verify the as-predicted intriguing electronic and electrical properties in the dislocated layered materials, including but not limited to, the current-generated magnetic field[65], Weyl fermions[66], chiraltronics[94] and non-trivial topologically protected modes[5]. Meanwhile, the axial screw dislocation spirally threads the vdW layers and possesses a covalent-bonded dislocation core, which can improve the electron conductivity along the cross-plane direction[95]. Also, the modulation of electron conductivity together with thermal conductivity along the cross-plane direction can be attractive for thermoelectrics. Third, exotic mechanical behaviors are worth extensive research interests in not only practical mechanical performance[69] but also fundamental mechanical laws[70].

## 6. SUMMARY

This review article summarizes the exciting progress in the SDD growth of layered materials, especially spiral nanoplates and twisted nanowires, as well as the proposed underlying mechanisms. We show the peculiar optical, electrical, thermoelectric, mechanical and catalytic properties in these dislocated systems, due to the unique symmetry. We also introduce the research of dislocations in the monolayer and bilayer 2D materials. Such advances in the dislocated layered materials will deepen our understanding of the material properties tailored by layer stacking symmetry, and further give rise to a wide range of state-of-the-art applications including optoelectronics, spintronics, valleytronics, and so on.

ACKNOWLEDGMENTS

R.C., J.C., S.G., and J.Y. acknowledge the support of Bakar Fellows Program at University of California, Berkeley.

REFERENCES

1. Hirth, J. P., Lothe, J. & Mura, T. Theory of Dislocations (2nd ed.). *Journal of Applied Mechanics* **50**, 476-477, doi:10.1115/1.3167075 (1983).
2. Burton, W., Cabrera, N. & Frank, F. Role of dislocations in crystal growth. *Nature* **163**, 398-399 (1949).
3. Anderson, P. M., Hirth, J. P. & Lothe, J. *Theory of dislocations*. (Cambridge University Press, 2017).
4. Szot, K., Speier, W., Bihlmayer, G. & Waser, R. Switching the electrical resistance of individual dislocations in single-crystalline SrTiO 3. *Nature materials* **5**, 312-320 (2006).
5. Ran, Y., Zhang, Y. & Vishwanath, A. One-dimensional topologically protected modes in topological insulators with lattice dislocations. *Nature Physics* **5**, 298-303, doi:10.1038/nphys1220 (2009).
6. Kim, S. I. *et al.* Dense dislocation arrays embedded in grain boundaries for high-performance bulk thermoelectrics. *Science* **348**, 109-114 (2015).
7. Sugiyama, I. *et al.* Ferromagnetic dislocations in antiferromagnetic NiO. *Nature nanotechnology* **8**, 266 (2013).
8. Gao, P. *et al.* Atomic-scale measurement of flexoelectric polarization at SrTiO 3 dislocations. *Physical review letters* **120**, 267601 (2018).
9. Geim, A. K. & Grigorieva, I. V. Van der Waals heterostructures. *Nature* **499**, 419-425 (2013).
10. Miró, P., Audiffred, M. & Heine, T. An atlas of two-dimensional materials. *Chemical Society Reviews* **43**, 6537-6554 (2014).
11. Xia, F., Wang, H., Xiao, D., Dubey, M. & Ramasubramaniam, A. Two-dimensional material nanophotonics. *Nature Photonics* **8**, 899 (2014).
12. Zhang, L. *et al.* Three-dimensional spirals of atomic layered MoS2. *Nano letters* **14**, 6418-6423 (2014).
13. Shearer, M. J. *et al.* Complex and noncentrosymmetric stacking of layered metal dichalcogenide materials created by screw dislocations. *Journal of the American Chemical Society* **139**, 3496-3504 (2017).
14. Suzuki, R. *et al.* Valley-dependent spin polarization in bulk MoS 2 with broken inversion symmetry. *Nature nanotechnology* **9**, 611 (2014).
15. Liu, Y. *et al.* Helical van der Waals crystals with discretized Eshelby twist. *Nature* **570**, 358-362, doi:10.1038/s41586-019-1308-y (2019).
16. Sutter, P., Wimer, S. & Sutter, E. Chiral twisted van der Waals nanowires. *Nature* **570**, 354-357 (2019).
17. Ly, T. H. *et al.* Vertically conductive MoS2 spiral pyramid. *Advanced Materials* **28**, 7723-7728 (2016).
18. Sarma, P. V. *et al.* Electrocatalysis on Edge-Rich Spiral WS2 for Hydrogen Evolution. *ACS Nano* **13**, 10448-10455, doi:10.1021/acsnano.9b04250 (2019).
19. Liu, L. *et al.* Bottom-up growth of homogeneous Moiré superlattices in bismuth oxychloride spiral nanosheets. *Nature Communications* **10**, 4472, doi:10.1038/s41467-019-12347-7 (2019).
20. Burton, W.-K., Cabrera, N. & Frank, F. The growth of crystals and the equilibrium structure of their surfaces. *Philosophical Transactions of the Royal Society of London. Series A, Mathematical and Physical Sciences* **243**, 299-358 (1951).

| | |
|---|---|
| 21 | Woodruff, D. How does your crystal grow? A commentary on Burton, Cabrera and Frank (1951)'The growth of crystals and the equilibrium structure of their surfaces'. *Philosophical Transactions of the Royal Society A: Mathematical, Physical and Engineering Sciences* **373**, 20140230 (2015). |
| 22 | Meng, F., Morin, S. A., Forticaux, A. & Jin, S. Screw Dislocation Driven Growth of Nanomaterials. *Accounts of Chemical Research* **46**, 1616-1626, doi:10.1021/ar400003q (2013). |
| 23 | Fan, X. *et al.* Controllable Growth and Formation Mechanisms of Dislocated WS2 Spirals. *Nano Letters* **18**, 3885-3892, doi:10.1021/acs.nanolett.8b01210 (2018). |
| 24 | Nie, Y. *et al.* Dislocation driven spiral and non-spiral growth in layered chalcogenides. *Nanoscale* **10**, 15023-15034 (2018). |
| 25 | Morin, S. A. & Jin, S. Screw Dislocation-Driven Epitaxial Solution Growth of ZnO Nanowires Seeded by Dislocations in GaN Substrates. *Nano Letters* **10**, 3459-3463, doi:10.1021/nl1015409 (2010). |
| 26 | Huang, X., Chumlyakov, Y. I. & Ramirez, A. G. Defect-driven synthesis of self-assembled single crystal titanium nanowires via electrochemistry. *Nanotechnology* **23**, 125601, doi:10.1088/0957-4484/23/12/125601 (2012). |
| 27 | Zhong, X., Shtukenberg, A. G., Hueckel, T., Kahr, B. & Ward, M. D. Screw Dislocation Generation by Inclusions in Molecular Crystals. *Crystal Growth & Design* **18**, 318-323, doi:10.1021/acs.cgd.7b01292 (2018). |
| 28 | O'Brien, M. *et al.* Mapping of Low-Frequency Raman Modes in CVD-Grown Transition Metal Dichalcogenides: Layer Number, Stacking Orientation and Resonant Effects. *Scientific Reports* **6**, 19476, doi:10.1038/srep19476 (2016). |
| 29 | Wu, J. *et al.* Spiral growth of SnSe2 crystals by chemical vapor deposition. *Advanced Materials Interfaces* **3**, 1600383 (2016). |
| 30 | Fan, X. *et al.* Broken symmetry induced strong nonlinear optical effects in spiral WS2 nanosheets. *ACS nano* **11**, 4892-4898 (2017). |
| 31 | Barman, P. K., Sarma, P. V., Shaijumon, M. & Kini, R. High degree of circular polarization in WS 2 spiral nanostructures induced by broken symmetry. *Scientific reports* **9**, 1-7 (2019). |
| 32 | Zhang, X. *et al.* Bandgap closing at the screw dislocations of WS2 spirals. *arXiv preprint arXiv:1906.07076* (2019). |
| 33 | Fan, X. *et al.* Mechanism of Extreme Optical Nonlinearities in Spiral WS2 above the Bandgap. *Nano Letters* **20**, 2667-2673 (2020). |
| 34 | Zhuang, A. *et al.* Screw-Dislocation-Driven Bidirectional Spiral Growth of Bi2Se3 Nanoplates. *Angewandte Chemie International Edition* **53**, 6425-6429 (2014). |
| 35 | Morin, S. A., Forticaux, A., Bierman, M. J. & Jin, S. Screw Dislocation-Driven Growth of Two-Dimensional Nanoplates. *Nano Letters* **11**, 4449-4455, doi:10.1021/nl202689m (2011). |
| 36 | Liu, L. *et al.* Bottom-up growth of homogeneous Moiré superlattices in bismuth oxychloride spiral nanosheets. *Nature communications* **10**, 1-10 (2019). |
| 37 | Karma, A. & Plapp, M. Spiral Surface Growth without Desorption. *Physical Review Letters* **81**, 4444-4447, doi:10.1103/PhysRevLett.81.4444 (1998). |
| 38 | Chen, L. *et al.* Screw-dislocation-driven growth of two-dimensional few-layer and pyramid-like WSe2 by sulfur-assisted chemical vapor deposition. *Acs Nano* **8**, 11543-11551 (2014). |
| 39 | Cao, Y. *et al.* Correlated insulator behaviour at half-filling in magic-angle graphene superlattices. *Nature* (2018). |
| 40 | Cao, Y. *et al.* Unconventional superconductivity in magic-angle graphene superlattices. *Nature* **556**, 43-50, doi:10.1038/nature26160 (2018). |
| 41 | Naik, M. H. & Jain, M. Ultraflatbands and Shear Solitons in Moiré Patterns of Twisted Bilayer Transition Metal Dichalcogenides. *Physical Review Letters* **121**, 266401, doi:10.1103/PhysRevLett.121.266401 (2018). |
| 42 | Jin, C. *et al.* Observation of moiré excitons in WSe 2/WS 2 heterostructure superlattices. *Nature*, 1 (2019). |
| 43 | Sharpe, A. L. *et al.* Emergent ferromagnetism near three-quarters filling in twisted bilayer graphene. *Science* **365**, 605-608, doi:10.1126/science.aaw3780 (2019). |
| 44 | Kennes, D. M., Xian, L., Claassen, M. & Rubio, A. One-dimensional flat bands in twisted bilayer germanium selenide. *Nature Communications* **11**, 1124, doi:10.1038/s41467-020-14947-0 (2020). |
| 45 | Eshelby, J. Screw dislocations in thin rods. *Journal of Applied Physics* **24**, 176-179 (1953). |
| 46 | Eshelby, J. The twist in a crystal whisker containing a dislocation. *Philosophical Magazine* **3**, 440-447 (1958). |
| 47 | Bierman, M. J., Lau, Y. A., Kvit, A. V., Schmitt, A. L. & Jin, S. Dislocation-driven nanowire growth and Eshelby twist. *Science* **320**, 1060-1063 (2008). |
| 48 | Zhu, J. *et al.* Formation of chiral branched nanowires by the Eshelby Twist. *Nature nanotechnology* **3**, 477 (2008). |
| 49 | Wu, H., Meng, F., Li, L., Jin, S. & Zheng, G. Dislocation-driven CdS and CdSe nanowire growth. *ACS nano* **6**, 4461-4468 (2012). |
| 50 | Jin, S., Bierman, M. J. & Morin, S. A. A new twist on nanowire formation: Screw-dislocation-driven growth of nanowires and nanotubes. *The Journal of Physical Chemistry Letters* **1**, 1472-1480 (2010). |
| 51 | Tizei, L. *et al.* Enhanced Eshelby twist on thin wurtzite InP nanowires and measurement of local crystal rotation. *Physical review letters* **107**, 195503 (2011). |
| 52 | Fang, Z. *et al.* Chemically modulating the twist rate of helical van der Waals crystals. *Chemistry of Materials* (2019). |
| 53 | Li, Y. *et al.* Probing symmetry properties of few-layer $MoS_2$ and h-BN by optical second-harmonic generation. *Nano letters* **13**, 3329-3333 (2013). |
| 54 | Schaibley, J. R. *et al.* Valleytronics in 2D materials. *Nature Reviews Materials* **1**, 16055, doi:10.1038/natrevmats.2016.55 (2016). |
| 55 | Mak, K. F., He, K., Shan, J. & Heinz, T. F. Control of valley polarization in monolayer $MoS_2$ by optical helicity. *Nature Nanotechnology* **7**, 494-498, doi:10.1038/nnano.2012.96 (2012). |
| 56 | Zhu, Z. Y., Cheng, Y. C. & Schwingenschlögl, U. Giant spin-orbit-induced spin splitting in two-dimensional transition-metal dichalcogenide semiconductors. *Physical Review B* **84**, 153402, doi:10.1103/PhysRevB.84.153402 (2011). |
| 57 | Cheiwchanchamnangij, T. & Lambrecht, W. R. L. Quasiparticle band structure calculation of monolayer, bilayer, and bulk $MoS_2$. *Physical Review B* **85**, 205302, doi:10.1103/PhysRevB.85.205302 (2012). |
| 58 | Xiao, D., Liu, G.-B., Feng, W., Xu, X. & Yao, W. Coupled Spin and Valley Physics in Monolayers of $MoS_2$ and Other Group-VI Dichalcogenides. *Physical Review Letters* **108**, 196802, doi:10.1103/PhysRevLett.108.196802 (2012). |
| 59 | Sarma, P. V., Patil, P. D., Barman, P. K., Kini, R. N. & Shaijumon, M. M. Controllable growth of few-layer spiral $WS_2$. *RSC Advances* **6**, 376-382, doi:10.1039/C5RA23020A (2016). |
| 60 | Lee, C. *et al.* Anomalous Lattice Vibrations of Single- and Few-Layer $MoS_2$. *ACS Nano* **4**, 2695-2700, doi:10.1021/nn1003937 (2010). |
| 61 | Puretzky, A. A. *et al.* Low-Frequency Raman Fingerprints of Two-Dimensional Metal Dichalcogenide Layer Stacking Configurations. *ACS Nano* **9**, 6333-6342, doi:10.1021/acsnano.5b01884 (2015). |
| 62 | Zhan, H., Zhang, G., Yang, C. & Gu, Y. Graphene Helicoid: Distinct Properties Promote Application of Graphene Related Materials in Thermal Management. *The Journal of Physical Chemistry C* **122**, 7605-7612, doi:10.1021/acs.jpcc.8b00868 (2018). |
| 63 | Mak, K. F., Lee, C., Hone, J., Shan, J. & Heinz, T. F. Atomically Thin $MoS_2$: A New Direct-Gap Semiconductor. *Physical Review Letters* **105**, 136805, doi:10.1103/PhysRevLett.105.136805 (2010). |
| 64 | Sarma, P. V., Patil, P. D., Barman, P. K., Kini, R. N. & Shaijumon, M. M. Controllable growth of few-layer spiral WS 2. *RSC advances* **6**, 376-382 (2016). |

65  Xu, F., Yu, H., Sadrzadeh, A. & Yakobson, B. I. Riemann surfaces of carbon as graphene nanosolenoids. *Nano letters* **16**, 34-39 (2016).
66  Wu, F., Zhang, R.-X. & Das Sarma, S. Three-dimensional topological twistronics. *Physical Review Research* **2**, 022010, doi:10.1103/PhysRevResearch.2.022010 (2020).
67  Zhou, X. *et al.* Ultrathin SnSe2 Flakes Grown by Chemical Vapor Deposition for High-Performance Photodetectors. *Advanced Materials* **27**, 8035-8041, doi:10.1002/adma.201503873 (2015).
68  Yu, P. *et al.* Fast Photoresponse from 1T Tin Diselenide Atomic Layers. *Advanced Functional Materials* **26**, 137-145, doi:10.1002/adfm.201503789 (2016).
69  Zhang, N., Carrez, P. & Shahsavari, R. Screw-dislocation-induced strengthening–toughening mechanisms in complex layered materials: The case study of tobermorite. *ACS applied materials & interfaces* **9**, 1496-1506 (2017).
70  Zhan, H., Zhang, G., Yang, C. & Gu, Y. Breakdown of Hooke's law at the nanoscale–2D material-based nanosprings. *Nanoscale* **10**, 18961-18968 (2018).
71  Björkman, T., Gulans, A., Krasheninnikov, A. V. & Nieminen, R. M. van der Waals bonding in layered compounds from advanced density-functional first-principles calculations. *Physical review letters* **108**, 235502 (2012).
72  Björkman, T., Gulans, A., Krasheninnikov, A. & Nieminen, R. Are we van der Waals ready? *Journal of Physics: Condensed Matter* **24**, 424218 (2012).
73  Hashimoto, A., Suenaga, K., Gloter, A., Urita, K. & Iijima, S. Direct evidence for atomic defects in graphene layers. *Nature* **430**, 870-873 (2004).
74  Gibb, A. L. *et al.* Atomic resolution imaging of grain boundary defects in monolayer chemical vapor deposition-grown hexagonal boron nitride. *Journal of the American Chemical Society* **135**, 6758-6761 (2013).
75  Van Der Zande, A. M. *et al.* Grains and grain boundaries in highly crystalline monolayer molybdenum disulphide. *Nature materials* **12**, 554-561 (2013).
76  Zhou, W. *et al.* Intrinsic structural defects in monolayer molybdenum disulfide. *Nano letters* **13**, 2615-2622 (2013).
77  Najmaei, S. *et al.* Vapour phase growth and grain boundary structure of molybdenum disulphide atomic layers. *Nature materials* **12**, 754-759 (2013).
78  Lahiri, J., Lin, Y., Bozkurt, P., Oleynik, I. I. & Batzill, M. An extended defect in graphene as a metallic wire. *Nature nanotechnology* **5**, 326 (2010).
79  Huang, P. Y. *et al.* Grains and grain boundaries in single-layer graphene atomic patchwork quilts. *Nature* **469**, 389-392 (2011).
80  Yazyev, O. V. & Louie, S. G. Topological defects in graphene: Dislocations and grain boundaries. *Physical Review B* **81**, 195420 (2010).
81  Liu, Y., Zou, X. & Yakobson, B. I. Dislocations and grain boundaries in two-dimensional boron nitride. *ACS nano* **6**, 7053-7058 (2012).
82  Zou, X., Liu, Y. & Yakobson, B. I. Predicting dislocations and grain boundaries in two-dimensional metal-disulfides from the first principles. *Nano letters* **13**, 253-258 (2013).
83  Alden, J. S. *et al.* Strain solitons and topological defects in bilayer graphene. *Proceedings of the National Academy of Sciences* **110**, 11256-11260 (2013).
84  Butz, B. *et al.* Dislocations in bilayer graphene. *Nature* **505**, 533-537 (2014).
85  Weston, A. *et al.* Atomic reconstruction in twisted bilayers of transition metal dichalcogenides. *Nature Nanotechnology*, 1-6 (2020).
86  Kisslinger, F. *et al.* Linear magnetoresistance in mosaic-like bilayer graphene. *Nature Physics* **11**, 650-653, doi:10.1038/nphys3368 (2015).
87  Chen, G. *et al.* Signatures of tunable superconductivity in a trilayer graphene moiré superlattice. *Nature* **572**, 215-219, doi:10.1038/s41586-019-1393-y (2019).
88  Chen, G. *et al.* Tunable correlated Chern insulator and ferromagnetism in a moiré superlattice. *Nature* **579**, 56-61, doi:10.1038/s41586-020-2049-7 (2020).
89  Seyler, K. L. *et al.* Signatures of moiré-trapped valley excitons in MoSe2/WSe2 heterobilayers. *Nature* **567**, 66-70, doi:10.1038/s41586-019-0957-1 (2019).

90  Jin, C. *et al.* Observation of moiré excitons in WSe2/WS2 heterostructure superlattices. *Nature* **567**, 76-80, doi:10.1038/s41586-019-0976-y (2019).
91  Hu, G. *et al.* Topological polaritons and photonic magic angles in twisted α-MoO3 bilayers. *Nature* **582**, 209-213, doi:10.1038/s41586-020-2359-9 (2020).
92  Engheta, N. & Pelet, P. Modes in chirowaveguides. *Optics Letters* **14**, 593-595 (1989).
93  Wang, X. *et al.* Current-driven magnetization switching in a van der Waals ferromagnet Fe3GeTe2. *Science advances* **5**, eaaw8904 (2019).
94  Atanasov, V. & Saxena, A. Helicoidal graphene nanoribbons: Chiraltronics. *Physical Review B* **92**, 035440 (2015).
95  Lee, S. *et al.* Anomalously low electronic thermal conductivity in metallic vanadium dioxide. *Science* **355**, 371-374 (2017).